\documentclass[a4paper,twocolumn,11pt,accepted=2025-06-20]{quantumarticle}
\pdfoutput=1
\usepackage{amsmath,amsfonts,amssymb,amsthm,bbm,graphicx,enumerate,times}
\usepackage{mathtools}
\usepackage[usenames,dvipsnames]{color}
\usepackage{hyperref}
\usepackage{comment}
\usepackage{graphicx}
\usepackage{float}
\usepackage{etoolbox}
\usepackage{accents}

\usepackage{array}
\usepackage[table]{xcolor}
\usepackage{xcolor}
\usepackage{dsfont}

\newcommand{\ii}{\mathrm{i}}

\newcommand{\tr}{\mathrm{tr}}

\newcommand{\m}{\mathrm{\gamma}}

\definecolor{darkgreen}{RGB}{0,200,100}

\begin{document}

\title{Critical spin models from holographic disorder}

\author{Dimitris Saraidaris}
\affiliation{Department of Physics, Freie Universit\"at Berlin, 14195 Berlin, Germany}
\email{d.saraidaris@fu-berlin.de}
\author{Alexander Jahn}
\affiliation{Department of Physics, Freie Universit\"at Berlin, 14195 Berlin, Germany}
\orcid{0000-0002-7142-0059}
\email{a.jahn@fu-berlin.de}

\begin{abstract}
    Discrete models of holographic dualities, typically modeled by tensor networks on hyperbolic tilings, produce quantum states with a characteristic quasiperiodic disorder not present in continuum holography.
    In this work, we study the behavior of XXZ spin chains with such symmetries, showing that lessons learned from previous non-interacting (matchgate) tensor networks generalize to more generic Hamiltonians under holographic disorder: While the disorder breaks translation invariance, site-averaged correlations and entanglement of the disorder-free critical phase are preserved at a plateau of nonzero disorder even at large system sizes.
    In particular, we show numerically that the entanglement entropy curves in this disordered phase follow the expected scaling of a conformal field theory (CFT) in the continuum limit. This property is shown to be non-generic for other types of quasiperiodic disorder, only appearing when our boundary disorder ansatz is described by a ``dual'' bulk hyperbolic tiling. Our results therefore suggest the existence of a whole class of critical phases whose symmetries are derived from models of discrete holography.

\end{abstract}

\maketitle

\section{Introduction}

Studying the holographic principle through the lens of quantum information theory has proven to be fruitful for both the field of quantum information and high-energy physics. In particular, within the context of the anti-de Sitter/conformal field theory (AdS/CFT) correspondence \cite{Witten1998,Maldacena1999}, holographic descriptions of quantum information concepts such as entanglement entropy \cite{Ryu:2006bv,Rangamani:2016dms}, continuum \cite{Almheiri:2014lwa} and tensor-network quantum error correction \cite{Pastawski:2015qua,Jahn2021}, and quantum cryptography \cite{Apel2024} have been found. 
One convenient approach to studying quantum information features of holographic systems is to use finite-dimensional toy models, in particular tensor networks with a hyperbolic lattice geometry \cite{Swingle:2009bg}. 
Such tensor networks are typically constructed as planar $2$-dimensional discretizations of time-slices of AdS$_{2+1}$ space-time, with open indices at the boundary of a disk-shaped region representing a discretized CFT state.
Under a regular discretization of the AdS bulk, these boundary states exhibit quasi-periodic symmetries on the boundary \cite{Boyle:2018uiv,Jahn:2019mbb,Jahn:2020ukq,Basteiro2022, Basteiro2023}, leading to the question which specific finite-dimensional boundary theories can be described by such tensor network models. 
The properties of a class of efficiently contractible \emph{matchgate tensor networks} \cite{valiant2002quantum,cai2007theory,Bravyi2008} can be explicitly computed on large hyperbolic tilings, reproducing the average properties of the critical Ising CFT with central charge $c=1/2$ \cite{Jahn2019}.
In this setting, the boundary Hamiltonian (i.e., the \emph{parent Hamiltonian} whose ground state coincides with the tensor network state) has been explicitly constructed \cite{Jahn2022_boundary}.
The resulting nearest-neighbor Hamiltonians obey quasiperiodicity on several length scales and can be well-approximated by the analytical \emph{multi-scale quasicrystal ansatz} (MQA), which we will describe below in detail. However, the matchgate tensor constraint of these studies restricted the boundary theory to one of free fermions, remote from the interacting boundary theories found in AdS/CFT.

In this work, we overcome this limitation by studying general boundary theories with MQA symmetries beyond its use as a phenomenological model in the hyperbolic tensor network setup of Ref.\ \cite{Jahn2022_boundary}.
First, we study the MQA symmetries for the $\{3,7\}$ hyperbolic tiling, describing how these symmetries translate into a sequence of couplings. With these couplings we generate disordered Gaussian Hamiltonians and study their critical behavior. Generalizing this analysis, we also study disordered non-Gaussian Hamiltonians via a matrix product state (MPS) ansatz for the boundary, thus extending beyond the Gaussian regime probed by matchgate tensor networks.
We then further generalize to a broad analysis of the MQA for other hyperbolic lattices as well as generic quasiperiodic sequences with and without the MQA.
In a nutshell, we investigate the robustness of criticality against quasiperiodic disorder in the following settings:
\begin{enumerate}
\setlength\itemsep{0cm}
    \item In the full parameter space of disordered nearest-neighbor Hamiltonians generated by the MQA for hyperbolic tilings.
    \item Using Hamiltonians that describe interacting theories, i.e., no longer have Gaussian fermionic ground states. 
    \item Generating MQA-disordered systems from arbitrary quasiperiodic sequences, including those not derived from the symmetries of a hyperbolic bulk geometry.
\end{enumerate}
We find that disordered critical phases with a CFT continuum limit appear generically for those types of the MQA that can be expressed in terms of a ``holographically dual'' bulk tiling.
Conversely, Hamiltonians with other types of quasiperiodic symmetries related to hyperbolic tilings, e.g.\ those symmetries derived from only a single layer of the bulk, are shown to exhibit other critical and non-critical phases without such a continuum limit.

\section{Results}

\subsection{Multi-scale quasiperiodic disorder}

The multi-scale quasicrystal ansatz (MQA) is a particular class of disordered $(1+1)$-dimensional spin or fermionic chains with nearest-neighbor interactions \cite{Jahn2022_boundary}. It was initially introduced as a disordered version of the transverse-field Ising Hamiltonian
\begin{align}
\label{EQ_H_ISING_DISORDERED}
    H_\text{I} = \frac{\ii}{2}\sum_{k=1}^{2L} J_k\, \m_k \m_{k+1} \ ,
\end{align}
written in terms of $2L$ Majorana modes $\m_k$ with $\{\m_j,\m_k\} = 2\delta_{j,k}$, and a set $\{J_k\}$ of local couplings. 
The standard translation-invariant form of the Ising model corresponds to setting all odd and even couplings $J_{k}$ to constants $J/2$ and $h$, respectively.
In this work we focus on periodic boundary conditions and identify $\m_{2L+1} \equiv -\m_1$ for consistency with the usual Ising model definition in the spin picture.

In an MQA, the couplings are defined from two conditions. The first determines the odd couplings $J_{2k+1}$, corresponding to couplings of Majorana modes within the same physical site, while the even couplings occur between Majorana modes on two neighboring sites. 
The former are determined from the latter, simply being set to the average of the couplings with adjacent sites:
\begin{align}
    J_{2k+1}=\frac{J_{2k}+J_{2k+2}}{2} \ .
\end{align}
The second condition of the MQA determines the even couplings $J_{2k}$ from a \emph{letter inflation rule}. Previous disorder constructions with quasiperiodic symmetry used couplings that alternate between two values $j_a$ and $j_b$ according to a quasiperiodic two-letter sequence \cite{Igloi_2007,Juhasz2007}, which is scaled up to larger sequences in each inflation step by replacing single letters with longer subsequences.
For example, the quasiperiodic \emph{Fibonacci sequence} emerges from iteratively applying the inflation rule $a \mapsto ab$, $b \mapsto a$.
The MQA, however, goes one step further: It superimposes the quasiperiodicity from \emph{all} steps of such an inflation rule by multiplying the couplings $j_a,j_b$ associated with each letter across iteration steps to form the physical coupling constants $J_k$.
This mimics a critical renormalization-group transformation in which self-similar disorder is introduced at every length scale, as in holographic tensor network constructions \cite{Boyle:2018uiv,Jahn:2019mbb,Jahn:2020ukq}.

\begin{figure*}[t]
\includegraphics[width=0.95\linewidth]{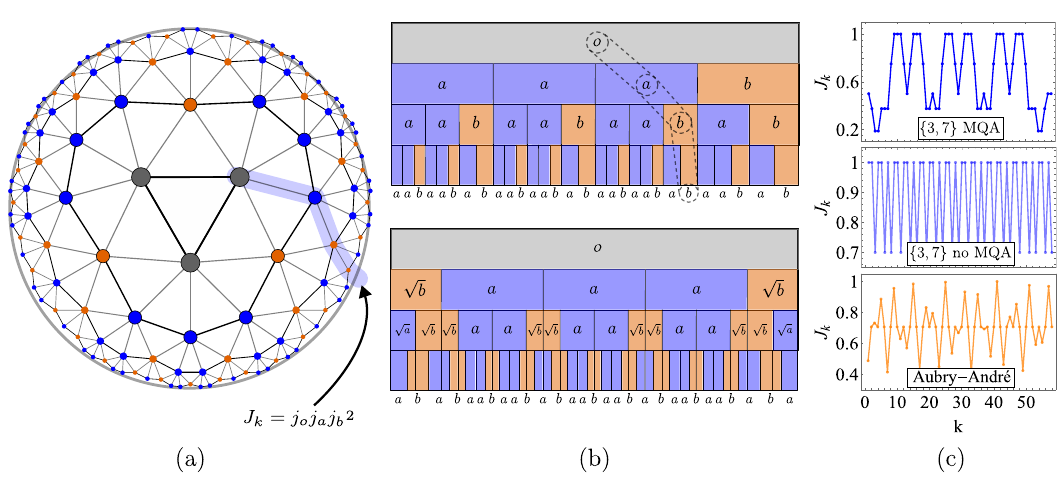}
\caption{(a): The first three inflation steps of the $\{3,7\}$ hyperbolic lattice in terms of $o$, $a$ and $b$ vertices (gray, blue and orange). Shaded in blue, we track a path along bulk vertices from the center to a given boundary vertex $k$, the product of couplings along which yields the coupling constant $J_k$ in the multi-scale quasicrystal ansatz (MQA). 
(b): (\textit{upper panel}) Visualization of the MQA for the first three inflation step. The dashed lines denote the bulk path shaded in (a). (\textit{lower panel}) Visualization of the MQA for the first three inflation steps with symmetrization.
(c): We show the coupling constant sequences for various cases, for all of which the disorder strength is the same. (\textit{upper panel}) The coupling constant sequence $J_k$ of the 3rd inflation step of the $\{3,7\}$ hyperbolic lattice after the application of the MQA. 
(\textit{middle panel}) The coupling constant sequence $J_k$ of the 3rd inflation step of the $\{3,7\}$ hyperbolic lattice without the application of the MQA. The resulting sequence is exactly the two-letter sequence determined by the corresponding inflation rule. 
(\textit{lower panel}) The coupling constant sequence $J_k$ generated by the quasiperiodic function $j_k=1+ D\cdot\cos(2\pi\frac{1+\sqrt{5}}{2}\cdot k )$ where the parameters have been choosen such that the disorder strength is the same as for the two sequences above.
}
  \label{fig:MQA}
\end{figure*}

Letter inflation rules appear naturally when describing layers of \emph{regular hyperbolic tilings}, i.e., tessellations of the hyperbolic disk by regular $p$-gons. 
These can be classified  by the Schl{\"a}fli symbols $\{p, q\}$, where $p$ denotes the number of edges of each tile, and $q$ is the number of such tiles that share the same vertex.
Hyperbolic triangular tilings are then denoted by $\{3,q\}$ with $q \geq 7$, the $q=7$ case of which is  pictured in Fig.~\ref{fig:MQA}(a).
While the bulk is regular, its boundary is quasiperiodic: Cutting off the tiling after $n$ layers of tiles (starting from a single central tile), the adjacency of open vertices varies along the boundary. Denoting vertices with two, three, and four adjacent edges (up to the cutoff) as $o$, $a$, and $b$, respectively, adding another layer to the tiling is equivalent to applying the inflation rule
\begin{align}
\sigma_{\{3,7\}} = 
\begin{cases}
    o \mapsto aaab \\
    a \mapsto aab \\
    b \mapsto ab
\end{cases}
\end{align}
to a sequence of letters denoting the previous boundary vertices. 
In Fig.~\ref{fig:MQA}(a), we begin with three $o$ vertices at the center (sequence $ooo$), shaded in gray. Starting from these vertices, we build the lattice step by step, following the inflation rule above, producing a sequence of only $a$ (blue) and $b$ vertices (orange) in the sequence $aaabaaabaaab$.
As we repeat this procedure, we produce a sequence of blue and orange vertices that alternates quasiperiodically (in addition to the initial $\mathbb{Z}_3$ symmetry). 
This symmetry also appears in tensor network models built from such geometries \cite{Boyle:2018uiv,Jahn:2019mbb,Jahn:2020ukq}: Identifying each tile with a $p$-leg tensor and contracting pairs of legs between neighboring tiles/tensors up to the cutoff, one would expect this geometric boundary symmetry to be reflected in the states produced by the tensor network, assuming that the tensors are chosen homogeneously and not break the symmetries of the tiling.
This assumption includes random tensor networks on such geometries \cite{Hayden:2016cfa}, where sample averaging will smooth out any deviations from the lattice symmetries, as well as holographic codes such as those introduced in Ref.\ \cite{Pastawski:2015qua}, whose code states also exhibit MQA symmetries \cite{Jahn:2020ukq}.
A previous analysis utilized \emph{matchgate tensor networks}, based on tensors with an efficiently computable free-fermion representation, to fill the bulk of the $\{3,7\}$ hyperbolic lattice, respecting its symmetries \cite{Jahn2019}. This structure led to boundary states that show a polynomial decay of site-averaged correlations, indicative of critical behavior and following the continuum CFT expectation for the critical Ising model.
Generally, we expect arbitrary disorder to interfere with critical behavior; the boundary states of hyperbolic tensor networks, however, do preserve criticality, and they allow for approximate local parent Hamiltonians of the form \eqref{EQ_H_ISING_DISORDERED} with varying couplings $J_k$ that follow an MQA \cite{Jahn2022_boundary,Jahn:2020ukq}. 

Consider the example of the $\{3,7\}$ hyperbolic tiling in Fig.~\ref{fig:MQA}(a): 
Here we associate ``elementary'' couplings $j_a$ and $j_b$ to each vertex type $a$ and $b$, and construct the physical couplings $J_k$ in Eq.\ \eqref{EQ_H_ISING_DISORDERED}, associated with modes on the tiling boundary, by multiplying all the elementary couplings across the bulk layers.
Keeping track of all vertices deep in the bulk is natural from a holographic perspective, as it is expected that the boundary encodes information at all levels of depth in the bulk, rather than merely those that are close to the boundary. 
There is an ambiguity in choosing this path through previous layers, and the naive construction (letter $x$ precedes $y$ if it is included in the inflation rule of $x$) may lead to a sequence that breaks the symmetries of the tiling, as shown in Fig.~\ref{fig:MQA}(b). This can be rectified by symmetrizing the inflation rules by splitting the letter $b$, for example $a \to \sqrt{b} a a \sqrt{b}$, where each $\sqrt{b}$ still corresponds to a coupling $j_b$.
We may also associate a coupling $j_o$ with the starting vertices $o$, which appears only once in each sequence and therefore acts merely as an overall normalization constant. Similarly, rescaling $j_a$ and $j_b$ by the same factor only changes the normalization, so that the eigenstates of $H$ only depend on the ratio $r = j_b/j_a$, resulting in a single free parameter of the $\{3,7\}$ MQA.
In the matchgate tensor network result of Ref.\ \cite{Jahn2022_boundary}, the $\{3,7\}$ tiling led to a ratio $r \approx 0.526$ depending on the geometric irregularity of the tiling boundary at a finite cutoff. 
Tilings with higher curvature, i.e., $\{p,q\}$ tilings with larger $p$ and $q$, exhibit a more irregular boundary and hence more disorder for the same value of $r$, so they are expected to exhibit disordered Ising-like MQA phases at values of $r$ closer to $1$.

Though the MQA was previously applied to a disordered Ising model \eqref{EQ_H_ISING_DISORDERED}, which is non-interacting and therefore exactly solvable with Gaussian techniques, it can also be generalized to other disordered Hamiltonians with nearest-neighbor couplings.
In particular, we consider the disordered anti-ferromagnetic Heisenberg model described by the Hamiltonian
\begin{align}\label{eq:Heisenberg}
    H_\text{H} = \sum_k^{L-1} J_k \big(\textbf{S}^x_k\textbf{S}^x_{k+1} + \textbf{S}^y_k\textbf{S}^y_{k+1} + \Delta_0\, \textbf{S}^z_k\textbf{S}^z_{k+1}  \big) \ ,
\end{align}
where $\textbf{S}_i$ are the spin-$\frac{1}{2}$ operators. 
Upon setting the anisotropy parameter $\Delta_0 = 0$, the model becomes non-interacting and can be decomposed into two copies of the disordered Ising Hamiltonian \eqref{EQ_H_ISING_DISORDERED}. 
For constant couplings $J_k=J$, the theory becomes critical for $0 \leq \Delta_0 \leq 1$. The three cases $\Delta_0=0$, $0<\Delta_0<1$ and $\Delta_0=1$ are called the Heisenberg XX, XXZ, and XXX model, respectively\footnote{The XX model is itself a special case of the XY model, in which the $\textbf{S}^x_k \textbf{S}^x_{k+1}$ and $\textbf{S}^y_k \textbf{S}^y_{k+1}$ terms can have different prefactors.}. While the first two cases obey a U(1) symmetry, there is a symmetry transition at $\Delta_0=1$ leading to the SU(2)-symmetric XXX model.

\subsection{Criticality in Gaussian MQA}

We now study the effect of applying the disorder couplings $J_k$ from the $\{3,7\}$ MQA for an arbitrary ratio $r=j_b/j_a$. Addressing the first two questions in the introduction, we consider how criticality depends on $r$ and whether it remains stable when turning on interactions.


One-dimensional uniform spin systems, like the Ising or XXZ-Heisenberg model with constant $J_k$ for all sites $k$, show critical behaviour characterized by polynomial correlation decay. This is also the expected behavior for a conformal field theory (CFT), which typically appears as the continuum limit of these spin chain models. 
Another generic feature of CFT ground states in $1{+}1$ dimensions is a logarithmic scaling of the entanglement entropy $S_A=-\tr[\rho_A\log\rho_A]$ \cite{Holzhey:1994we}. Including finite size corrections to a periodic system of length $L$, the entanglement entropy for a subregion $A$ of length $\ell$ follows the form \cite{Calabrese:2004eu,Calabrese2009}
\begin{align}
\label{eq:CalabreseCardy}
    S_A &= \frac{c}{3}\log\left( \frac{L}{\pi \epsilon}\sin\frac{\pi\ell}{L}\right) + \kappa  \\
    &\sim \frac{c}{3} \log\frac{\ell}{\epsilon} \quad\text{ for } \ell \ll L \ .
\end{align}
where $c$ is the central charge of the corresponding CFT, $\epsilon \ll \ell$ is a sufficiently small lattice constant and $\kappa$ is CFT-dependent. 
\begin{figure*}[ht]
\includegraphics[width=1.0\linewidth]{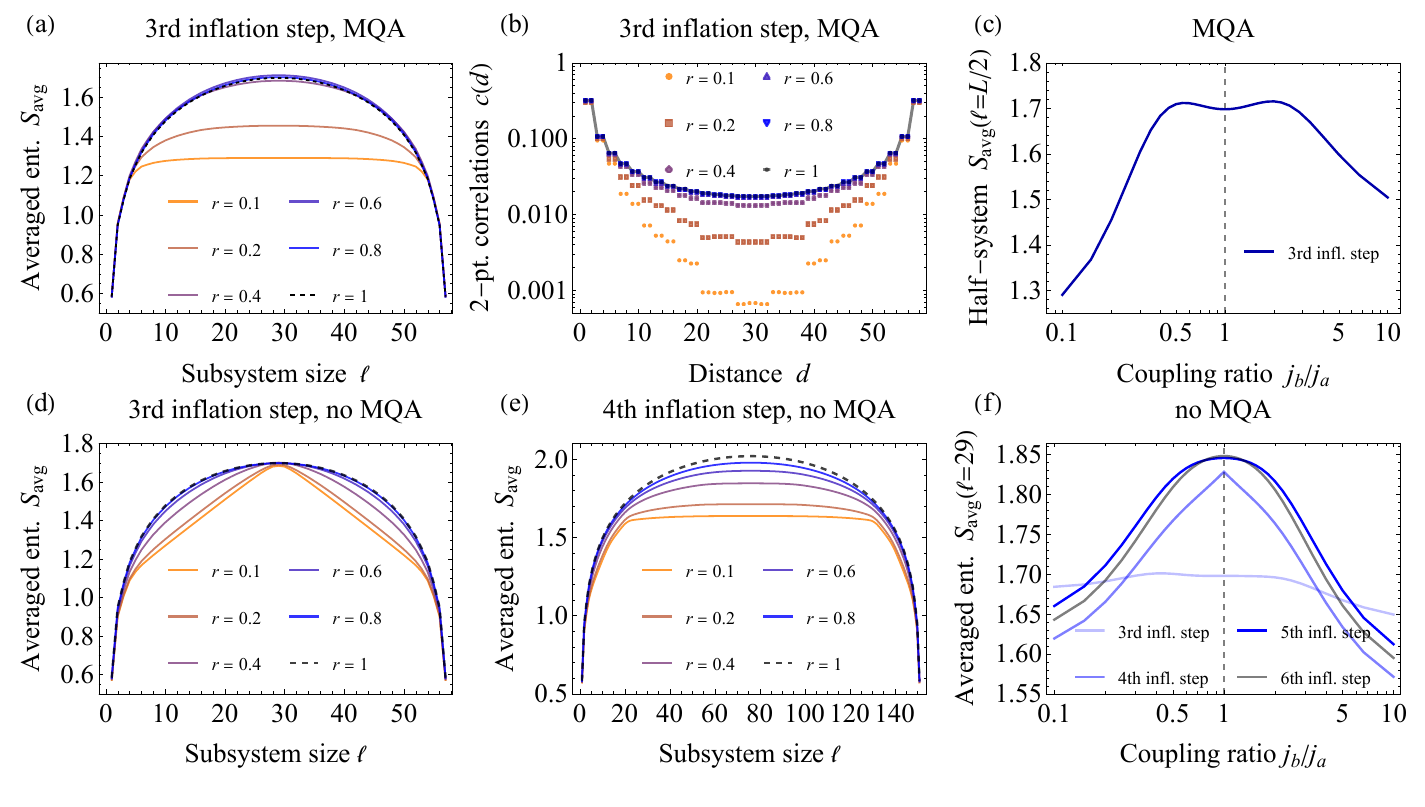}
\caption{(a): Site-averaged entanglement entropy scaling over subsystem size $\ell$ for the 3rd inflation step of the $\{3,7\}$ hyperbolic lattice after applying the MQA. Each colored curve corresponds to a different ratio $r=j_b/j_a$, thus different disorder. The dashed black curves correspond to the critical behavior of the XX-Heisenberg model, without disorder, in perfect agreement with Eq.~\eqref{eq:CalabreseCardy}. (b):  Averaged 2-point correlation decay over distance $d$, for different ratios $r$. (c): Averaged entanglement entropy of $\ell=L/2=29$-spin subsystems over different ratios $r$ (disorder). (d): Averaged entanglement entropy scaling over subsystem size $\ell$, for the 3rd inflation step of the $\{3,7\}$ hyperbolic lattice without applying the MQA. (e):  Averaged entanglement entropy scaling over subsystem size $\ell$, for the 4th inflation step of the $\{3,7\}$ hyperbolic lattice without applying the MQA. (f): Averaged entanglement entropy of $\ell=29$-spin subsystems over different ratios $r$ for the 3rd, 4th, 5th and 6th inflation step of the $\{3,7\}$ hyperbolic lattice without applying the MQA.}
\label{fig:373}
\end{figure*}
Here we study disordered spin systems breaking translation invariance, thus requiring site-averaging relevant quantities.
Defining $S_j(\ell)$ as the entanglement entropy of a compact interval $A = [j,\, (j+\ell-1) \operatorname{mod} L]$, we can write the average over a system with periodic boundary conditions as
\begin{equation}
\label{eq:averaged_entanglement_entropy}
    S_\text{avg}(\ell) = \frac{1}{L} \sum_{j=1}^L S_j(\ell) \ .
\end{equation}
For the Gaussian model ($\Delta_0=0$) we calculate the entanglement entropy of different subsystems using the covariance matrix formulation \cite{Surace2022} (see Appendix \ref{APP_COV}). This formulation cannot be applied for non-Gaussian cases, such as the XXZ-Heisenberg model ($\Delta_0\neq 0$), where tensor network approximations are needed, as we will discuss later. 

In Fig.~\ref{fig:373}(a), we plot  $S_\text{avg}(\ell)$ for the ground state of a disordered Hamiltonian with coupling constants provided by the 3rd inflation step of the $\{3,7\}$ hyperbolic lattice (58 sites with 116 Majorana modes).  The parameter $r=j_b/j_a$ is the ratio between the values of the  blue and orange vertices, which controls the disorder strength: The further the ratio $r$ is from $r_{\mathrm{crit}} = 1$, the stronger the disorder. For strong disorder, the systems shows a saturation in the entanglement entropy scaling after a specific length scale, indicative of a gapped system expected to arise in that limit. 
However, for ratios $r\gtrsim 0.5$ the system shows entanglement entropy scaling in perfect agreement with the uniform, translation-invariant case ($r_{\mathrm{crit}}=1$) following the Calabrese-Cardy ansatz \eqref{eq:CalabreseCardy} for a CFT. 
This transition with $r$ becomes apparent in Fig.~\ref{fig:373}(c), where we plot the averaged entanglement entropy $S_\text{avg}(\ell)$ of subsystems of size $\ell=L/2$ for different ratios $r$. There is a distinct plateau around the region {$0.5 < r < 2$} where all the disordered models have entanglement entropy values close to the one for the non-disordered model. 
This plateau indicates that moderate MQA disorder preserves criticality, with the specific ratio  $r\sim 0.526$ appearing in matchgate tensor networks in Ref.\ \cite{Jahn2022_boundary} describing only one point in this critical phase. Interestingly, this point is close to a local maximum in entanglement entropy at $r_{\mathrm{max}}\approx 0.55$, as shown in Fig.~\ref{fig:373}(b). At this point, $S_\text{avg}$ even becomes slightly larger than at the translation-invariant point $r=1$. Note that such a local maximum with the same $S_\text{avg}$ also exists around $r \approx 2$, suggesting a symmetry $r \leftrightarrow \frac{1}{r}$ in the critical disordered phase.
The appearance of such an approximate symmetry at small $|r-1|$ is likely a result of a dependence of critical behavior on the overall disorder strength, which exhibits a similar symmetry; for example, if we measure disorder strength in terms of the standard deviation of the sequence of couplings $J_k$, we find the same approximate $r\leftrightarrow\frac{1}{r}$ symmetry.
We can compare the entanglement scaling under MQA disorder to the case of aperiodic disorder with a 2-letter sequence $J_k \in (j_a, j_b)$  taken from the letters at only the last inflation step that generated a corresponding MQA (e.g.\ the last row of Fig.\ \ref{fig:MQA}(b)).
In our analysis, we will use the shorthand \emph{non-MQA} to refer to this simpler type of quasiperiodic disorder.
In this case, the $r \to 0$ and $r \to \infty$ limits can be described by the \emph{strong disorder renormalization group} (SDRG),
as local pairs of sites coupled by either $j_a \gg j_b$ or $j_b \gg j_a$ can be iteratively replaced by singlets.
This singlet-based renormalization process leads to a characteristic $S_\text{avg}$ scaling in linear segments with a logarithmic envelope \cite{Juhasz2007}.
Figs.~\ref{fig:373}(d) and (e) show the gradual transition from the translation-invariant to the disordered case for this simpler type of disorder, at $n=3$ and $n=4$ inflation steps. Note that the logarithmic envelope of the entanglement entropy for small subsystem sizes $\ell$ has a smaller prefactor than the translation-invariant value of $1/3$, showing a transition to a different phase in the large-disorder limit (this deviation becomes increasingly visible at larger $n$).
Unlike the MQA, this simpler aperiodic disorder therefore does not have a stable critical phase at small disorder: As we show in Fig.~\ref{fig:373}(f), any deviation from the translation-invariant case $r=1$ results in an immediate decline of subsystem entanglement entropy.
In CFT language, this disorder corresponds to a relevant operator that drives the system to another phase (the critical SDRG phase). 

In addition to entanglement entropy scaling, we can study the critical behaviour of the disordered boundary states by analyzing the averaged correlation decay of the ground states of the specific disordered Hamiltonians. 
For Hamiltonians of the Gaussian form \eqref{EQ_H_ISING_DISORDERED}, it can be computed as
\begin{align}
    c(d) = \frac{1}{2L}\sum_{j=1}^{2L} | \Gamma_{j, j+d} |
    \label{eq:correlation_decay}
\end{align}
where $\Gamma_{i,j}$ is the ground state covariance matrix in the Majorana picture (see App.\ \ref{APP_COV} for details). Here we average over all the matrix elements that correspond to pairs of Majorana modes at distance $d$.

For $\Delta_0=0$, the Hamiltonian of the Heisenberg model \eqref{eq:Heisenberg} becomes Gaussian, thus of the form of \eqref{EQ_H_ISING_DISORDERED}, so we can calculate the correlation decay as described above. The disordered cases of the free-fermion XX-Heisenberg model show exponential decay  for $r<0.5$ and $\ell<L/2$, while the cases for $r\gtrsim 0.5$ show polynomial decay which is close to the uniform critical model, as seen in Fig.~\ref{fig:373}(b). Again, there is a regime of Gaussian disordered cases, derived from the boundary of the hyperbolic tiling by translating it into a sequence of coupling constants, that shows critical behavior. This is also in perfect agreement with the entanglement entropy results above.

\begin{figure*}[t]
\centering
\includegraphics[width=0.98\linewidth]{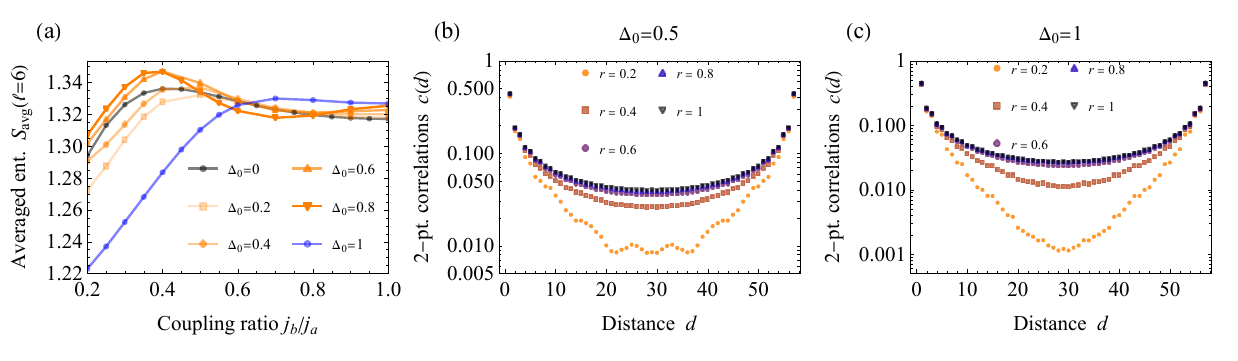}
\caption{(a): Averaged entanglement entropy of subsystems of size $\ell=6$ over different ratios $r$. We consider the ground states of the $\{3,7\}$ MQA-disordered Heisenberg model (3rd inflation step) for $\Delta_0=0$ (black line), the non-Gaussian cases for $\Delta_0=\{0.2, 0.4, 0.6, 0.8\}$ (orange lines) and the isotropic SU$(2)$ model for $\Delta_0=1$ (blue line). (b): Spin-spin correlations decay of the ground state of the $\{3,7\}$ MQA-disordered Heisenberg model for $\Delta_0=0.5$. (c): Spin-spin correlations decay of the ground state of the $\{3,7\}$ MQA-disordered Heisenberg model for $\Delta_0=1$ (SU$(2)$-symmetric XXZ-model).}
  \label{fig:ent_nongaussian}
\end{figure*}

\subsection{Criticality in Non-Gaussian MQA}

Up to this point we have only considered ground states of Gaussian systems, i.e., those that can be mapped to a free-fermion Hamiltonian. This was also the limit of previous matchgate tensor studies and analytical arguments in Ref.\ \cite{Jahn2022_boundary}. This raised the question whether MQA disorder only preserves criticality in the Gaussian setting. To resolve it, we now consider MQA-disordered ground states of the XXZ Hamiltonian \eqref{eq:Heisenberg} in the non-Gaussian regime $\Delta_0\neq 0$.

As free-fermion techniques no longer apply, we turn to a numerical tensor network approach to search for the ground state of the Hamiltonian and determine its properties. A widely established method for general 1-dimensional Hamiltonians is given by the \emph{Density Matrix Renormalization Group} (DMRG) \cite{PhysRevLett.69.2863,SchollwoeckDMRG}. For the following analysis, we apply the DMRG with bond dimension $D=3000$ and periodic boundary conditions to approximate the ground state of the disordered XXZ Hamiltonian expressed in a matrix product state (MPS) form. We can calculate the entanglement entropy of relatively small sub-systems ($\ell\leq 8$) by forming the corresponding reduced density matrix out of two copies of the ground state MPS after having traced out the rest of the system. Then we calculate the averaged entanglement entropy as defined in \eqref{eq:averaged_entanglement_entropy}.

In Fig.~\ref{fig:ent_nongaussian}, we plot the averaged entanglement entropy for the Gaussian model ($\Delta_0=0$) with a black line. This is qualitatively the same result as in Fig.~\ref{fig:373}(c), but for subsystem size $\ell=6$ and only for $j_b / j_a <1$. In different shades of orange, we draw the entanglement entropy scaling for the ground states of the non-Gaussian Hamiltonians for different values of the anisotropy constant $\Delta_0$. For all values of $\Delta_0$ studied, we note that the systems show close-to-critical behavior for a range of ratios in the disorder regime. If we consider the case for $\Delta_0=1$ (blue line), where now the system is SU$(2)$-symmetric, the entanglement entropy scaling shows similar behavior, with the difference that the width of the criticality plateau is shorter. This difference might be present due to the symmetry transition from U(1) to SU(2) at $\Delta_0=1$, which has been a topic of recent interest \cite{Lastres2024, Rylands2024}.

Additionally, we calculate the averaged spin-spin correlation decay, defined for the non-Gaussian Hamiltonians ($\Delta_0\neq 0$) as:
\begin{equation}
\label{eq:spinspincorrelations}
    c(d) = \frac{1}{L}\sum_{j=1}^{L} \langle \mathbf{S}_j\cdot \mathbf{S}_{j+d} \rangle
\end{equation}

In Fig.~\ref{fig:ent_nongaussian}(b, c) we show the spin-spin correlation decay for two non-Gaussian Hamiltonians, in particular for $\Delta_0=0.5$ and $\Delta_0=1$, respectively. In both cases, we notice that there exist coupling ratios away from $r=1$ at which we see polynomial correlation decay that is very similar to the critical case at $r=1$. This is in perfect agreement with the averaged entanglement entropy results in Fig.~\ref{fig:ent_nongaussian}(a), discussed above, providing one more piece of evidence that this critical disordered regime is robust even when non-Gaussianity is considered.

Though these results were produced with a 1-dimensional MPS, they suggest that similar states can be produced by more general 2-dimensional hyperbolic tensor networks on a $\{3,7\}$ geometry, for which MQA-type disorder is an expected feature. This would require foregoing the computational advantages of using matchgate tensors, which only produce Gaussian states, but could be useful to construct models of holographic dualities, where boundary states are typically in the strong-coupling regime.

\begin{figure*}[ht]
\centering
\includegraphics[width=\textwidth]{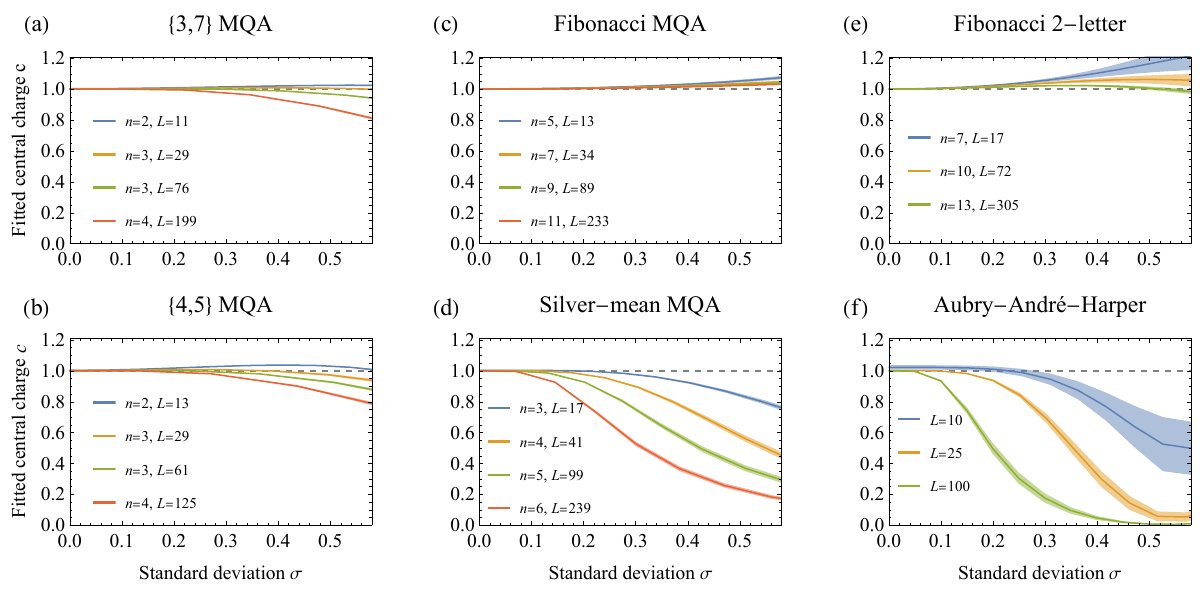}
\caption{
Stability of effective central charge $c$ under disorder, fitted according to the entanglement entropy scaling \eqref{eq:CalabreseCardy}, relative to the standard deviation $\sigma$ of XX model coupling sequences $J_k$. The disorder-free $c=1$ phase is denoted by a dashed line.
(a)-(b): For the MQA based on the $\{3,7\}$ and $\{4,5\}$ regular hyperbolic tilings after $n$ inflation steps with $L$ total sites, $c$ remains stable at small $\sigma$ but decays at large $\sigma$, indicating a breakdown of the critical phase.
(c)-(d): The Fibonacci and silver-mean MQA, which are not derived from hyperbolic inflation rules, show different behavior: The Fibonacci MQA remains stable at $c\approx 1$ even at large system sizes $L$ and disorder $\sigma$, while in the silver-mean MQA, $c$ decays quickly in both limits.
(e)-(f): Two examples of non-MQA quasiperiodic sequences: The Fibonacci 2-letter sequence shows decay only at large $L$ and $\sigma$, but is expected to reach an aperiodic singlet phase with $c=0.796$ at $\sigma\to\infty$, while the Aubry-Andr\'e-Harper model decays quickly towards $c\to 0$ as disorder is increased.
For all plots, the curve width represents the parameter error of each fit.
}
\label{fig:central_charge_fits}
\end{figure*}

\subsection{Non-MQA quasiperiodic disorder}

Having found numerical evidence that the behavior of MQA disorder leads to a critical phase for a range of coupling ratios and that this behavior is generic to both Gaussian and non-Gaussian critical systems, we now approach the third question stated in the Introduction: How does criticality depend on the specific quasiperiodic symmetries induced by an MQA derived from an inflation rule of a hyperbolic tiling, compared both to ``non-holographic'' MQA constructions from an inflation rule without such an associated bulk geometry and to other quasiperiodic disorder types.

\begin{table}
\scriptsize
\begin{center}
\setlength{\tabcolsep}{0pt}
\begin{tabular}{| c | c | c | c | c || c | c |}
 \hline
 \mbox{} & \hspace{2pt}$\{3,7\}$\hspace{2pt} & \hspace{3pt}$\{3,8\}$\hspace{3pt}  & \hspace{2pt}$\{4,5\}$\hspace{2pt} & \hspace{6pt}$\{6,4\}$\hspace{6pt} & \hspace{2pt}Fibonacci\hspace{2pt} & \hspace{2pt}Silver-mean\hspace{2pt} \\ 
 \hline
\rowcolor[RGB]{215, 215, 215} $o\rightarrow$ &$aaab$& $aaaab$ & - & - & - & - \\
 \hline
  \rowcolor[RGB]{155, 155, 225} $a\rightarrow$ & $aab$ & $aaab$ & $ababa$ & $aba$ &  $ab$ & $aba$\\  
 \hline
\rowcolor[RGB]{244, 176, 123} $b\rightarrow$ & $ab$ & $aab$ & $aba$ & $abaaaba$ &$a$ & $a$\\
\hline
\end{tabular}
\end{center}
\caption{We present the inflation rules for all the aperiodic two-letter sequences we have studied, both the hyperbolic and non-hyperbolic ones.}
\label{table:inflation_rules}
\end{table}

To compare the resulting critical and non-critical phases under various types of disorder, we use a best-fit of $S_\text{avg}$ over the whole range of subsystem sizes $\ell$ with respect to the Calabrese-Cardy formula \eqref{eq:CalabreseCardy} at different system sizes $L$ (an alternative analysis in terms of the behavior of $S_\text{avg}$ at fixed $\ell$ is presented in Appendix \ref{SEC_ALT_DISORDER}).
Fig.\ \ref{fig:central_charge_fits} shows the result of these fits relative to the standard deviation $\sigma$ of the coupling sequences $J_k$ of a (non-interacting) XX model Hamiltonian. 
In addition to the $\{3,7\}$ and $\{4,5\}$ MQA, we also show two examples of non-holographic MQA derived from the Fibonacci and silver-mean inflation rules (shown in Table \ref{table:inflation_rules}), as well as two other types of non-MQA  quasiperiodic disorder: Coupling sequences from the 2-letter Fibonacci sequence (i.e., the last layer of the corresponding MQA) and the \emph{Aubry-Andr\'e-Harper model} \cite{AubryAndre1980,Harper1955} with couplings
\begin{align}
    J_{2k} &= 1 + D\cdot\cos\left( 2\pi k\frac{1+\sqrt{5}}{2} \right) \ , \\ 
    J_{2k-1} &= J = \text{const} \ ,
\end{align}
where the disorder is controlled by a parameter $D$. We fix the odd couplings to a constant $J=\frac{2}{L} \sum_{k=1}^{L/2} J_{2k}$ corresponding to the mean of all even couplings, equivalent to a choice $J=1$ at system size $L\to\infty$.

\begin{figure}[tb]
\includegraphics[width=1\linewidth]{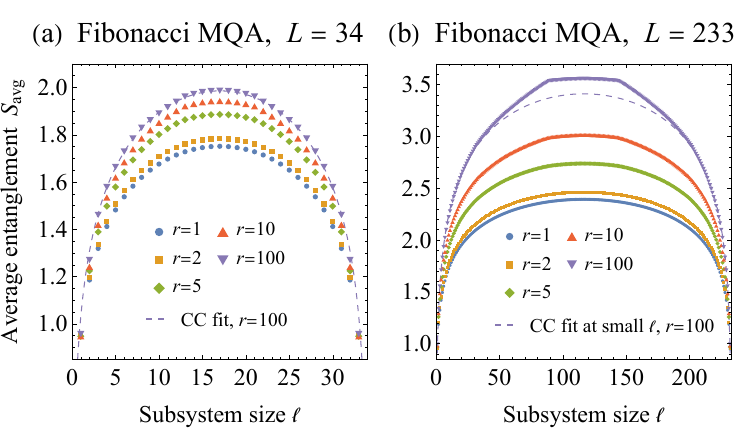}
\caption{
Averaged entanglement entropy $S_\text{avg}$ for the Fibonacci MQA at (a) $n=7$ and (b) $n=11$ inflation steps at different coupling ratios $r = j_b/j_a$. 
Unlike other MQA constructions, entanglement increases monotonically as disorder is increased.
At large $n$ and $r$, $S_\text{avg}$ also deviates from a Calabrese-Cardy (CC) fit \eqref{eq:CalabreseCardy}, exhibiting a segmented linear growth typical for an aperiodic singlet phase.
}
\label{fig:central_charge_fib}
\end{figure}

At $\sigma=0$, all of these models in Fig.\ \ref{fig:central_charge_fits} recover the effective central charge $c=1$ of the translation-invariant critical XX model.
Once a disorder with $\sigma>0$ is introduced, however, the three types of sequences (holographic + MQA, non-holographic + MQA, and quasiperiodic non-MQA) show qualitative differences.
We find that the $\{3,7\}$ and $\{4,5\}$ MQA both produce a fit with $c \approx 1$ (i.e., equivalent to the translation-invariant case) for a moderate range of disorder even at large $L$, consistent with the results for the $\{3,7\}$ tiling discussed above. And large $\sigma$, both systems transition to a gapped phase with $c \to 0$.
In contrast, the silver-mean MQA leads to a $c$ fit that monotonically decays to zero with $\sigma$ even for small system sizes $L$, with the decay rapidly increasing at larger $L$. This indicates the absence of any disordered critical phase in the continuum limit.
The Fibonacci MQA, however, does not show a decay of the fitted $c$ at all: At finite $L$, $c$ grows slowly with $\sigma$, while in the $L \to\infty$ limit it approaches a $\sigma$-independent $c=1$.
At first glance, this appears to indicate a plateau behavior at small disorder similar to the $\{p,q\}$ MQAs. However, a closer look at the actual entanglement entropy curves $S_\text{avg}(\ell)$ shows significant deviations from the translation-invariant case. 
As shown in Fig.\ \ref{fig:central_charge_fib}(a), it is still possible to fit the data points with a Calabrese-Cardy curve \eqref{eq:CalabreseCardy} at small system sizes $L$, however only with values of $\kappa$ and $c$ larger than the translation-invariant values (e.g.\ $\kappa\approx 1.03, c\approx0.4$ for $r=100$ compared to $\kappa\approx0.96,c=1/3$ at $r=1$).
Fig.\ \ref{fig:37vsTheWorld}(c) in the Appendix confirms the absence of an entanglement plateau at small disorder, further distinguishing it from the $\{p,q\}$ MQAs.
As we consider the Fibonacci MQA at larger system sizes in Fig.\ \ref{fig:central_charge_fib}(b), we find that $S_\text{avg}(\ell)$ starts to deviate from a smooth Calabrese-Cardy curve at large $r$, beginning to exhibit sections of linear growth with a logarithmic envelope, a characteristic feature of an aperiodic singlet phase described by the SDRG (compare Fig.~\ref{fig:373}(d)).
Such phases appear when most of the entanglement is mediated by (close to) maximally entangled pairs, which indeed appears to be the case here: Spin-spin correlations in the $r \to \infty$ limit become increasingly sparse but remain long-range. This limit strongly deviates from any other MQA we have studied, where the limit of an infinite coupling ratio $r$ always results in a gapped phase; here, we appear to find a transition to another critical phase.

Finally, the two quasiperiodic but non-MQA sequences considered, the Fibonacci 2-letter sequence and the Aubry-Andr\'e-Harper model, show wildy different behavior.
The former leads to an initially increasing $c$ at $\sigma>0$, but converges to a smaller value at large $L$ and $\sigma$. This behavior is well-understood: It corresponds to the strong-disorder limit of an aperiodic singlet phase with an effective central charge \cite{Juhasz2007}
\begin{equation}
    c_\text{Fib} = \left( 3- \frac{3}{\sqrt{5}} \right) \frac{\log 2}{\operatorname{arsinh} 2} \approx 0.796 \ .
\end{equation}
Again, this limit leads to an $S_\text{avg}(\ell)$ that increases in linear segments, as shown in Fig.~\ref{fig:373}(d).
Unlike the Fibonacci case, Aubry-Andr\'e-Harper disorder rapidly destabilizes the critical phase for all but very small systems; again, this indicates that there is no stable critical phase at small disorder. Note, however, that a disordered critical point does appear at \emph{large} disorder $D=2$, where the model undergoes a metal-insulator phase transition \cite{AubryAndre1980,jitomirskaya1999metal,Bu_2022}. However, the effective central charge at this critical point differs from the $c=1$ critical point of the XX model \cite{Goncalves2024}.

\begin{figure*}[t]
\centering
\includegraphics[width=0.98\linewidth]{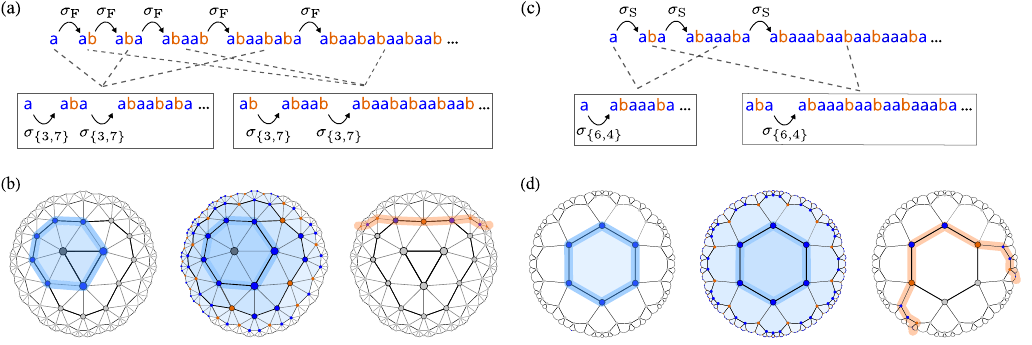}
\caption{(a): The Fibonacci sequence as a mixing of two $\{3,7\}$ aperiodic sequences. (b): A $\{3,7\}$ sequence starting from an initial letter $a$ can be represented on a $\{3,7\}$ hyperbolic lattice as a closed surface, which can cover the whole lattice, following the corresponding inflation steps (blue shaded). On the contrary, a $\{3,7\}$ sequence starting from initial letters $ab$ can be represented as a \textit{geodesic} on a $\{3,7\}$ hyperbolic lattice (orange shaded). (c) The silver-mean sequence as a mixing of two $\{6,4\}$ aperiodic sequences. (d): Starting with $a$, we can draw a closed surface on a $\{6,4\}$ hyperbolic lattice, which will fill the whole lattice, following the corresponding inflation steps (blue shaded). A sequence starting with $aba$, however, can be represented as a boundary-to-boundary cut on the $\{6,4\}$ hyperbolic lattice (orange shaded). }
  \label{fig:Fibo_Silver_hyperbolic}
\end{figure*}

\begin{figure}[tb]
\centering
\includegraphics[width=1\linewidth]{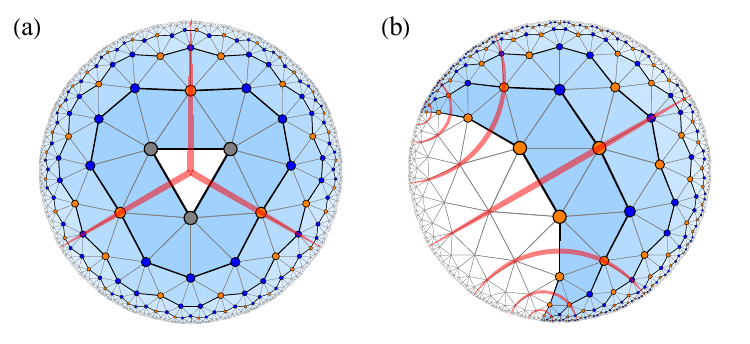}
\caption{
Inflation on the $\{3,7\}$ tiling, starting from (a) vertices around a single central tile with sequence $ooo$ and (b) an infinite strip of vertices with sequence $\overline{b}$. While the asymptotic boundary of (a) coincides with the whole boundary of the hyperbolic disk, (b) asymptotes to either of two parts of a boundary bipartition. Red curves denote the hyperbolic symmetry axes of each inflation.
}
  \label{fig:inflation_types}
\end{figure}

\subsection{Bulk geometry and holography}

As we have seen, a ``non-holographic'' MQA that is based on an inflation rule not describing a hyperbolic $\{p,q\}$ tiling appears unstable, in the continuum limit, under even small disorder. We will now explore the two such models considered above, the Fibonacci and silver-mean MQA, in more detail and provide a geometrical interpretation of their behavior.
Curiously, both inflation rules can be related to $\{p,q\}$ inflations under a doubling of the rule. For example, applying the Fibonacci rule twice leads to $a \mapsto ab \mapsto aba$ and $b \mapsto a \mapsto ab$, which is equivalent to the $\{3,7\}$ inflation rule under a shift by two letters. Similarly, a doubling of the silver-mean sequence can be shown to be equivalent to the inflation rule for a hyperbolic $\{6,4\}$ tiling. These equivalences are shown in Fig.~\ref{fig:Fibo_Silver_hyperbolic}, where we also visualize how repetitions of different initial sequences correspond to different bulk cuts in the hyperbolic tilings.
While this implies that both the Fibonacci and silver-mean sequences are asymptotically indistinguishable from the $\{3,7\}$ and $\{6,4\}$ sequences (i.e., for an arbitrarily large subsequence after sufficiently many inflation steps), the resulting MQA differ: For example, the same final sequence of the Fibonacci MQA will be constructed from twice as many layers as the corresponding $\{3,7\}$ MQA, with the even and odd layers each corresponding to an inflation of different bulk regions. 

We now give an interpretation to this behavior in terms of holographic symmetries. While the MQA is a general ansatz class that can be constructed from any inflation rule, those corresponding to the symmetries of regular hyperbolic tilings are endowed with the additional symmetries of the discretized hyperbolic (Poincar\'e) disk, of the so-called \emph{Fuchsian groups} \cite{Osborne:2017woa,Boettcher:2021njg}. This corresponds to a discrete version of the symmetries one finds in holographic dualities such as AdS/CFT, where the $SO(d,2)$ symmetries of $d{+}1$-dimensional anti-de Sitter (AdS) spacetime match the symmetries of a conformal field theory with $d$ spacetime dimensions \cite{Brown:1986nw}.  
Under a regular discretization of AdS time-slices, i.e., the hyperbolic disk, the spatial bulk symmetries that would otherwise match those of CFT ground states become reduced to a Fuchsian group, appearing on the boundary as a set of quasiperiodic symmetries \cite{Boyle:2018uiv,Jahn:2019mbb}. Theories whose ground states match these more general symmetries (which include the translation-invariant CFT ground states) have been proposed to belong to a new class of \emph{quasiperiodic conformal field theories} (qCFTs) \cite{Jahn:2020ukq}. At small disorder, the $\{p,q\}$ MQA XXZ models considered in this work are candidates for such qCFTs, as their symmetries, built from those of a $\{p,q\}$ regular hyperbolic tiling, are expected to match a bulk Fuchsian group. Following the logic above, these symmetries would therefore be a subset of the symmetries of ground states of the critical XXZ model with a $c=1$ continuum CFT limit. For the other MQA models, such as the Fibonacci and silver-mean MQA, such symmetries are not guaranteed. As shown in Fig.\ \ref{fig:Fibo_Silver_hyperbolic}, even and odd layers of both MQA are associated with different regions on the same $\{p,q\}$ tiling, on which bulk symmetries act differently. Specifically, the letter sequences describing the boundaries of these regions, both of which are periodically repeated, correspond to two different cuts through the hyperbolic disk: Closed curves of finite length centered around a single vertex or tile, leading to a rotational $\mathbb{Z}_p$ or $\mathbb{Z}_q$ symmetry, or infinitely long cuts between two boundary points with a translational $\mathbb{Z}_\infty$ symmetry (see Fig.\ \ref{fig:inflation_types}). A bulk transformation preserving one of these symmetries necessarily breaks the other, thus ensuring that the resulting MQA no longer respects qCFT symmetries.

\section{Discussion}

Understanding the structure of critical boundary theories in holographic bulk-boundary dualities with discretized geometries, such as tensor network models, is an essential step for understanding the relationship between discrete and continuum holography. This discretization necessarily leads to new boundary symmetries not present in the continuum case, suggesting the existence of new disordered critical many-body phases.
As we have shown, there exists strong numerical evidence that boundary symmetries generated by a bulk hyperbolic geometry -- analytically captured by the multiscale quasicrystal ansatz (MQA) -- lead to disordered critical phases whose entanglement entropy scaling is consistent with the continuum limit of a conformal field theory (CFT). These critical phases seem to be robust for a range of disorder values, appear in both interacting and non-interacting spin chains, and can be generated from different $\{p,q\}$ hyperbolic tilings.
The appearance of interacting disordered phases is of particular interest here, as they provide a discrete-holographic analogue of boundary theories in continuum holography that are typically considered in the strong coupling regime.
Crucially, the MQA only preserves the properties of the translation-invariant model required for a CFT continuum limit --- in particular, the same Calabrese-Cardy scaling \eqref{eq:CalabreseCardy} of entanglement entropy --- when it is given a quasiperiodic inflation sequence as an input that also corresponds to the symmetries of hyperbolic tiling layers:
As examples for non-hyperbolic MQA that break these symmetries, we saw that a silver-mean MQA does not have stable critical entanglement entropy scaling at increasing disorder (Fig.~\ref{fig:central_charge_fits}(d)) while a Fibonacci MQA shows immediate deviations from the translation-invariant entanglement entropy scaling with increasing disorder and a transition to a strongly-disordered phase in the infinite-disorder limit.
As MQA symmetries derived from hyperbolic tilings have previously appeared in tensor network models of holography, they can be properly identified as \emph{discrete-holographic boundary symmetries}.
Conversely, non-MQA quasiperiodic disorder, e.g.\ derived purely from the symmetries of the outermost layer of a hyperbolic tiling, leads to other critical (Fig.~\ref{fig:373}(d) and Fig.~\ref{fig:central_charge_fits}(e)) and non-critical (Fig.~\ref{fig:central_charge_fits}(f)) phases.
Hence, boundary theories of discrete-holographic models appear to achieve CFT-like critical behavior only when combining hyperbolic lattice symmetries with an MQA.
The hyperbolic tensor-network picture appears to suggest that the layers of such an MQA can be understood as a renormalization group transformation beyond the SDRG paradigm. Such an approach and its connection to (discretized) imaginary-time path integrals will be an avenue for future work.

Our numerical results further support the conjecture of Refs.\ \cite{Jahn:2020ukq,Jahn2022_boundary} that translation-invariant Hamiltonians with a CFT continuum limit retain their critical properties under a disorder derived from the symmetries of regular hyperbolic tilings, forming a new class of quasiperiodic conformal field theories (qCFT).
Intriguingly, we have found here that such behavior even appears without the explicit construction of a tensor network on such hyperbolic tilings, but that boundary theories with the proper (MQA) disorder exhibit qCFT-like critical phases.
Studying discrete holography thus appears to lead to distinct many-body phases that are not apparent in models of continuum holography. 
The precise mathematical relationship between these qCFTs and the CFTs emerging from the non-disordered continuum limit is an interesting question for further study.
As the MQA disorder contains contributions on all length scales and does not disappear under coarse-graining, it cannot be described by a CFT perturbation by an irrelevant operator; likewise, our work suggests that average correlations and entanglement of the non-perturbed phase are preserved, ruling out a relevant operator. It thus appears that a qCFT should be describable by a marginal perturbation of a continuum CFT.

Note that even though the MQA was derived from the symmetry of hyperbolic tensor network models, it is still possible to construct such tensor networks to produce non-MQA boundary states \cite{Basteiro2023}; however, such constructions then explicitly break the symmetries of the underlying tiling (e.g.\ by introducing a dependence on the radial direction) such that the qCFT symmetries are no longer upheld.
While holographic MQAs appear to exhibit universal critical properties, the non-holographic cases (such as the Fibonacci MQA) lead to other interesting phases that could be studied in greater detail in the future.

The tensor network models considered in this paper do not include bulk degrees of freedom, unlike models of \emph{holographic codes} \cite{Almheiri:2014lwa,Pastawski:2015qua,Jahn2021}. However, such tensor network codes produce boundary (code) states that also exhibit MQA disorder \cite{Jahn:2020ukq}, albeit with very sparse correlation functions that do not seem to lead to a smooth continuum theory. However, more sophisticated tensor-network models of AdS/CFT should be capable of combining (approximate) holographic codes that contain (quasiperiodic) CFTs in their code space under a suitable continuum limit. These boundary theories will likely resemble the MQA-disordered models studied here, as they will obey similar symmetries. General boundary states in these codes will also be non-Gaussian, though some basis states in holographic codes may still be describable as Gaussian states \cite{Jahn:2019nmz}.

Finally, a future challenge will be to implement and study actual physical systems with MQA disorder, for example in quantum simulators with trapped ions \cite{Joshi:2023rvd} or Rydberg atoms \cite{Yoshida2024}, with the goal of both studying the robustness of MQA criticality under physical noise as well as laying the groundwork for future implementations of discrete-holographic simulations.

\section*{Acknowledgments}
We wish to thank Lennart Bittel and Tommaso Guaita for fruitful discussions on Gaussian states, Jens Eisert for helpful comments on boundary conditions in critical spin chains, and Zolt\'an Zimbor\'as for relevant discussions on aperiodic disorder.
For all tensor network numerical simulations done in this paper, we used the QSpace tensor library
\cite{Weichselbaum2012, Weichselbaum2020,qspace4u} which can exploit general non-Abelian symmetries in tensor network algorithms. 
The authors are grateful for support from the Einstein Research Unit ``Perspectives of a quantum digital transformation''.

\bibliographystyle{quantum}
{

}

\newpage

\appendix

\section{Covariance matrix formalism and entanglement entropy calculation}
\label{APP_COV}

A fermionic Gaussian state \cite{Surace2022} can be defined as any state
\begin{align}
    \rho=\frac{e^{-\hat{H}}}{Z}
\end{align}
whose parent Hamiltonian $\hat{H}$ is a fermionic Gaussian Hamiltonian (i.e., containing no higher than quadratic terms in creation/annihilation operators).
The parent Hamiltonian contains all the information regarding such a state $\rho$. For any fermionic quadratic Hamiltonian, there is a fermionic transformation $\tilde{c}_k=Uc_k$ that diagonalizes it, thus a fermionic Gaussian state $\rho$ can be decomposed in terms of single-mode thermal states as follows: 
\begin{align}
    \rho= \bigotimes_{k=1}^N \frac{e^{-\epsilon_k(\tilde{c}_k^\dagger \tilde{c}_k - \tilde{c}_k\tilde{c}_k^\dagger)}}{Z_k} \ .
\end{align}
The above expression reveals that the fermionic Gaussian state $\rho$ is completely characterized by the occupation numbers $\langle \tilde{c}^\dagger_k \tilde{c}_k\rangle$ and as a consequence by the correlators $\langle c_i^\dagger c_j\rangle$ and $\langle c_i c_j\rangle$ after we apply the reverse transformation $c_k=U^\dagger \tilde{c}_k$.
It is convenient to study a Gaussian model in the Majorana fermion picture. In this case the Gaussian fermionic state $\rho$ is completely characterized by the covariance matrix
\begin{align}
    \Gamma_{j,k} = \frac{\ii}{2} \langle \psi | \gamma_j\gamma_k - \gamma_k\gamma_j | \psi \rangle = \frac{\ii}{2}\mathrm{Tr}\big(\rho [\gamma_j, \gamma_k]\big) \ ,
\end{align}
\noindent where the density matrix is thermal and $\gamma_k$ are the Majorana operators with $\{\gamma_j,\gamma_k\} = 2 \delta_{j,k}$. The covariance matrix $\Gamma$ is a real, skew-symmetric matrix, which makes its diagonalization practical: For any real, skew-symmetric $2N \times 2N$ matrix $M$, there exists a real special orthogonal matrix $O\in SO(2N)$, such that $M=O\tilde{M}O^\text{T}$, where $\tilde{M}$ is a block-diagonal matrix of the form
\begin{align}
    \Tilde{M} = OMO^\text{T} = \bigoplus_{j=1}^N \begin{pmatrix}
        0 & \mu_j \\
        -\mu_j & 0
    \end{pmatrix} \ ,
\end{align}
\noindent where $\mu_j\in \big[ -\frac{1}{2}, \frac{1}{2} \big]$. 
The eigenvalues $\epsilon_k$ of the Hamiltonian $H$ are then given by
\begin{align}
    \epsilon_k =\frac{1}{2} \ln \bigg( \frac{1+\mu_j}{\frac{1}{2}-\mu_j} \bigg) \ .
\end{align}
Having obtained the eigenvalues $\mu_j$, the von Neumann entanglement entropy is given by
\begin{align}
    S(\rho) = \sum_{k=1}^N \big[ \nu_k\ln (\nu_k) + (1-\nu_k)\ln (1-\nu_k) \big] \ ,
\end{align}
where $\nu_k=\frac{1}{2}-\mu_k$. The block-diagonal form of the covariance matrix makes the calculation of the entanglement entropy of subsystems of any length $\ell$ quite convenient.

\begin{figure}[t!]
\centering
\includegraphics[width=1\linewidth]{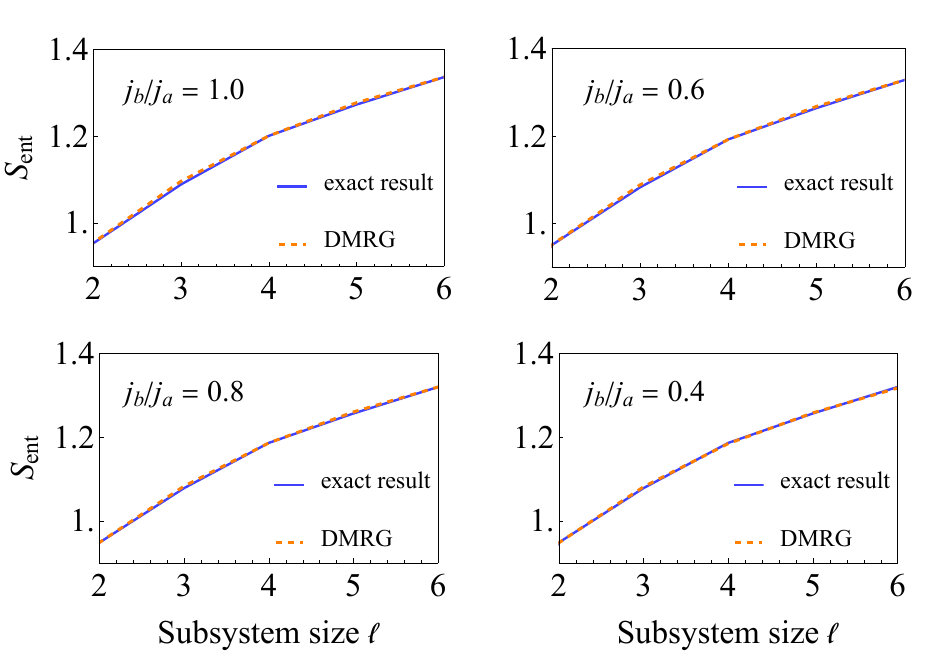}
\caption{We benchmark the DMRG ground state approximation with the exact result out of the covariance matrix formalism, by calculating their averaged entanglement entropy scaling over subsystem size for the Gaussian XX-Heisenberg model ($\Delta_0=0$) and disorder cases $r\in\{0.4, 0.6, 0.8, 1\}$.}
  \label{fig:benchmark}
\end{figure}

\begin{figure*}[t!]
\centering
\includegraphics[width=\linewidth]{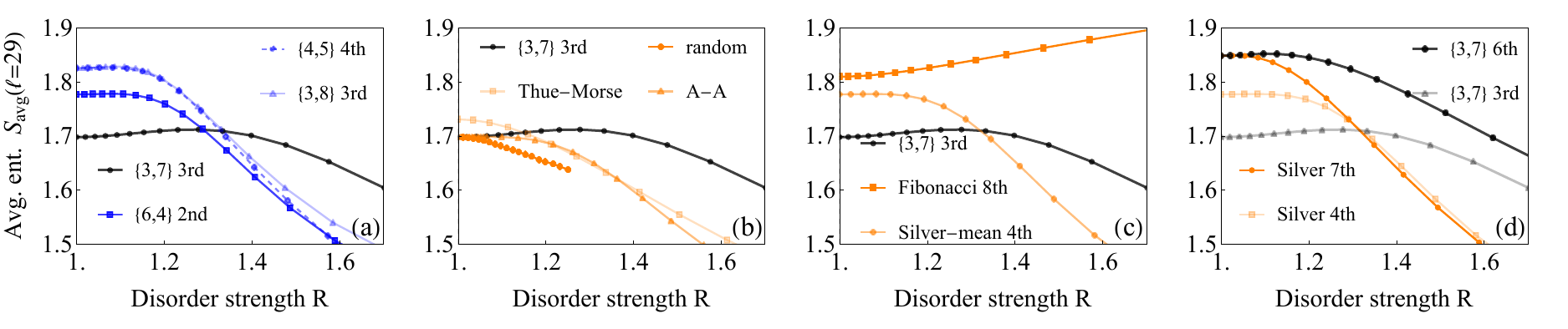}
\caption{Site-averaged entanglement entropy $S_\text{avg}$ for subsystems of size $\ell=29$ over disorder strength for (a): hyperbolic sequences and (b): non-hyperbolic aperiodic sequences. In both (a) and (b) we present the result of the $\{3,7\}$ hyperbolic lattice with a black line for comparison. (c): Averaged entanglement entropy for subsystems of size $\ell=29$ over disorder strength for the Fibonacci and Silver-mean aperiodic sequences. (d): We present the behavior of the averaged entanglement over disorder strength for the 6th inflation step of the $\{3,7\}$ hyperbolic sequence and the 7th inflation step of the Silver-mean sequences. Each represents the asymptotic behavior of the averaged entanglement entropy at a large number of inflation steps for these two cases.
}
  \label{fig:37vsTheWorld}
\end{figure*}

\section{Benchmarking DMRG}
\label{APP_BENCHMARK}

We can check the accuracy of the DMRG method we used for the disordered non-Gaussian models in our analysis above by applying it also to the Gaussian model. For the Gaussian model we can obtain the entanglement entropy exactly via the covariance matrix that describes the Gaussian ground state of each disordered Hamiltonian. For the same disordered Hamiltonian, we can find an approximation of the ground state via DMRG, and then calculate the entanglement entropy by manipulating the MPS ground state accordingly. 

In Fig.~\ref{fig:benchmark}, we plot the averaged entanglement entropy over the subsystem size for disordered Gaussian Hamiltonians, generated using the couplings of the 3rd inflation step of the $\{3,7\}$ hyperbolic lattice. As we can see, the DMRG method approximates the ground state of the Gaussian models with great accuracy for bond dimension $D=1024$.

\section{Alternative measure of disorder}
\label{SEC_ALT_DISORDER}

We consider the disorder dependence of site-averaged entanglement entropy $S_\text{avg}$ against various types of disorder to check the stability of a critical phase. 
Here we deviate from the analysis in the main text in two ways: First, we consider $S_\text{avg}(\ell)$ at a fixed subsystem length $\ell = 29$ (half of the $\{3,7\}$ boundary after three inflation steps). Second, we define a new disorder measure $R=\mathrm{mean}\bigg(\frac{\mathrm{max}(J_k,J_{k+1})}{\mathrm{min}(J_k,J_{k+1})}\bigg)$ that captures the deviation from translation invariance for arbitrary coupling sequences, superseding the ratio $r$ between bulk couplings in order to compare different kinds of disorder.

In Fig.~\ref{fig:37vsTheWorld}(a), we plot $S_\text{avg}(\ell=29)$ as a function of the disorder strength for the MQA derived from the inflation rules of the $\{3,7\}$, $\{6,4\}$, $\{4,5\}$, and $\{3,8\}$ hyperbolic lattices (see Table \ref{table:inflation_rules} for the specific inflation rules). For all these case of ``holographic'' disorder, we can distinguish plateaus of different widths that expand in the disordered regime. 
We contrast this behavior with non-MQA aperiodic sequences: In Fig.~\ref{fig:37vsTheWorld}(b), we plot $S_\text{avg}(\ell=29)$ for the Thue-Morse aperiodic sequence ($a\mapsto ab, b\mapsto ba$) as well as Aubry-Andr\'e couplings, for both of which it decreases monotonically with increasing $R$.
Interestingly, if we draw couplings randomly from a Gaussian distribution (also shown in Fig.~\ref{fig:37vsTheWorld}(b)), the entanglement entropy again decreases monotonically with $R$, without any signs of a disordered critical regime. Specifically, we drew couplings $J_i$ from a normal distribution with mean $\mu=10$ and variance $\sigma^2\in [0.25,3]$. Each data point corresponds to a fixed $\{\mu,\sigma^2\}$ with 2000 random samples, with their average disorder strength $R$ and entanglement $S_\text{avg}(\ell=L/2)$ being plotted.
These results show that the MQA plateau behavior is very non-generic for random or even quasiperiodic disorder. Of course, any single random sample may deviate sharply from the sample average, in special cases producing high entanglement even for strong disorder.

Finally, we compare the holographic $\{3,7\}$ ansatz to non-holographic MQA derived from the Fibonacci and silver-mean sequences, both for small (Fig.~\ref{fig:37vsTheWorld}(c)) and larger system sizes (Fig.~\ref{fig:37vsTheWorld}(d)). 
While the silver-mean MQA leads to a small entanglement plateau at small system size, this plateau shrinks in the scaling limit while the $\{3,7\}$ case remains more stable. The Fibonacci MQA, on the other hand, shows a peculiar behavior: The entanglement entropy \emph{increases} monotonically with increased disorder, with a minimum reached in the disorder-free, translation-invariant case. 
This differs drastically with what we observed for the holographic MQAs (or any other type of studied disorder), where adding large disorder always reduces entanglement and induces a transition to a gapped (non-critical) phase. 
These results are in agreement with the central charge fits presented in Fig.\ \ref{fig:central_charge_fits} of the main text, confirming that our observations are robust against different choices of disorder parameters.

\clearpage
\end{document}